Biocompatible L-Cysteine Capped MoS$_2$ Nanoflowers for Antibacterial Applications: Mechanistic Insights

Rupal Kaushik[1], Suvendu Nandi[2], Mahitosh mandal[2], Amar Nath Gupta[1]*

[1] Soft Matter and Biophysics Laboratory, Department of Physics, IIT Kharagpur, 721302, India

[2] Cancer Biology Laboratory, School of Medical Science and Technology, IIT Kharagpur, 721302, India

* Corresponding author, *Email address*: ang@phy.iitkgp.ac.in (Amar Nath Gupta)

**Abstract:**

The development of multi-drug-resistant bacterial infections seriously threatens public health, so efforts are needed to develop a new class of safe and effective antibacterial agents. Here, we report a bio-inspired synthesis of surface-modified MoS$_2$ Nanoflowers with L-cysteine (MoS$_2$-cys NFs) that show good colloidal stability in an aqueous medium. The FE-SEM and TEM data confirm the flower-like morphology and determine the size of NFs (537± 12 nm); the XRD data predicts the hexagonal crystal structure of the NFs. The XPS peaks confirm the formation of MoS$_2$ NFs with surface modification by L-cysteine. FTIR measurements also confirm the presence of L-cysteine in the NFs. The antibacterial activity of as-prepared MoS$_2$-cys NFs examined over gram-negative *Escherichia coli* and gram-positive *Staphylococcus aureus* bacteria shows inhibition of nearly 97% and 90%, respectively, at concentrations 250 μg/mL each after incubation of 6 hrs. The antibacterial mechanism is mainly attributed to the generation of oxidative stress, which can occur through both ROS-dependent and ROS-independent pathways. The ROS-dependent and ROS-independent oxidative stresses were measured using flow cytometry and fluorescence imaging using DCFH-DA staining, and Ellman's assay, respectively. Moreover, the toxicity studies of the MoS$_2$-cys NFs towards HFF cell lines suggested the high biocompatibility of the nanomaterial with a cell viability of nearly 90%. We report the intrinsic antibacterial efficiency of MoS$_2$-cys NFs without any external stimulus (light, H$_2$O$_2$, etc.), doping, or drug loading. Our study indicates that the proper surface modification can enhance the colloidal stability and antibacterial potency of MoS$_2$-based nanomaterials for further applications such as antibacterial coatings, water disinfection, and wound healing.

## 1. Introduction:

Microbial infections are one of the major threats to human health, accounting for millions of deaths worldwide[1]. Antibiotics have traditionally been used to treat these infections. However, overuse of these traditional antibiotics is leading to the development of antimicrobial resistance[2,3]. Therefore, there is an urgent need to design and develop novel antibacterial agents to address this problem. Recently, substantial efforts have been made to develop the potential nanomaterial-based antibiotics to counter multidrug-resistant infections and improve their antibacterial efficiency[4,5]. In this context, silver nanoparticles have been extensively studied and applied as antimicrobial agents for water purification, wound healing, and antibacterial coatings in the textile and food packaging industries[6]. Despite this, their potential as antibacterial drugs is constrained by their toxicity to the host[7]. Antibiotics based on gold nanoparticles have also demonstrated effectiveness against multidrug-resistant bacteria[8]. Beyond metal nanoparticles, two-dimensional (2-D) nanomaterials have also emerged as promising antibacterial agents in the last decade[9].

In this category, a 2-D nanomaterial Molybdenum disulfide (MoS$_2$), in particular, has earned significant attention due to its remarkable properties such as high catalytic activity, biocompatibility, low cost, ease of synthesis, and large surface area, hence its ability to communicate more with different biomolecules[10]. These properties have led to their use in various biomedical applications such as water disinfection, drug delivery, phototherapy, biosensing, antibacterial and anticancer agents, and wound healing[11,12]. The nanostructures have great potential as antibacterial agents, but their aggregation limits their biomedical applications. Yang et al. recently reported the stable dispersions of chemically exfoliated $MoS_2$ nanosheets (NSs) had shown nearly double antibacterial activity than the aggregated sheets due to their greater interaction abilities with the bacteria[13]. Hence, their study suggests that aggregating $MoS_2$ nanostructures in an aqueous medium decreases their antibacterial efficiency. Therefore, for $MoS_2$ to truly realize its full potential in biological applications, its surface must be appropriately modified to avoid aggregation and improve its solubility in aqueous media. Various reports suggest that the surface modification of nanoparticles with a nature motif may improve their interactions with biosystems, biocompatibility and increase their solubility in the aqueous medium[14,15].

L-cysteine, a hydrophilic sulfur-containing semi-essential amino acid, is extensively used to synthesize different cysteine-capped nanomaterials[16–19]. L-cysteine is a biocompatible, easy-to-obtain, environment-friendly, and inexpensive biomolecule. Moreover, its ability to coordinate with inorganic cations and metals via its three active functional groups (–$NH_2$, –COOH, and –SH) makes it an ideal capping agent and a greener sulfur source [20,21]. The cysteine-capped nanostructures increase the aqueous stability and anchor other molecules due to the free amino and carboxyl groups on their surfaces[21]. These studies make L-cysteine a suitable ligand for synthesizing nanomaterials for various applications. Recently, various studies have also reported the synthesis of cysteine-modified $MoS_2$ nanostructures for different biomedical applications[22–24]. Furthermore, the previous reports also suggest that L-cysteine exhibits antibacterial activity, due to which they have been applied on multiple fabrics for durable antibacterial properties[25,26]. Also, the distinct antibacterial activities of D-cysteine and L-cysteine on different bacterial strains have been reported[27].

Various studies have been conducted so far based on $MoS_2$ nanocomposites in order to enhance their antibacterial properties. The majority of these works are focussed on the use of thiolated surfactants and ligands for functionalization[28,29], introducing doping elements[30], loading drugs/antimicrobial peptides onto the nanomaterial[31,32], and combining it with other nanomaterials like silver, ZnO, $TiO_2$ etc[33–35]. Moreover, photothermal, photodynamic, photocatalytic properties, and peroxidase like activities of $MoS_2$ have led to development of synergistic external stimuli responsive antibacterial system with enhanced antibacterial efficiency of the $MoS_2$-based nanocomposites[36–40]. Zhao et al. reported the synergistic antibacterial activity of NIR-assisted streptomycin sulfate loaded PEG-$MoS_2$/reduced graphene oxide (PEG-$MoS_2$/rGO-SS) nanoflakes. The results reveals that, the antibacterial activity of PEG-$MoS_2$ alone is very feeble as compared to PEG-$MoS_2$/rGO-SS nanoflakes both with and without NIR treatment [37]. Hence, it was found in all aforementioned studies that $MoS_2$ nanostructures alone (without functionalisation, drug loading, or external stimuli) have shown very low antibacterial activity as compared to the whole nanocomposite.

Moreover, $MoS_2$ nanoflowers (NFs) are gaining attention due to their unique flower-like morphology consisting of many NSs stacked in a three-dimensional (3D) nanostructure to provide greater surface area and more active edges than their 2D NSs[12]. It is expected that due to more active sites, NFs will possess good antibacterial activity. Recently, polyethylene glycol (PEG) functionalized $MoS_2$ NFs have shown negligible antibacterial activity in absence of $H_2O_2$ and NIR irradiation[36]. To the best of our knowledge, research based on $MoS_2$ NFs is still in the infant stage for antibacterial applications. Very few reports are there, based on surface-modified $MoS_2$ NFs, exploring their intrinsic antibacterial properties without any drug loading or external probes. In addition to this, limited studies have reported the colloidal stability of as-synthesised nanostructures.

Hence, in this work, we synthesized L-cysteine-capped $MoS_2$ NFs using facile hydrothermal synthesis method. The as-synthesised $MoS_2$-cys NFs have shown good antibacterial activity, with 97 % and 90 % inhibition against *E. coli* (gram-negative) and *S. aureus* (gram-positive) bacteria, respectively, at 250 µg/mL and 6 hours of incubation. The $MoS_2$-cys NFs are highly stable in aqueous solution, as shown through microplate reader and DLS measurements. Moreover, as-synthesized $MoS_2$-cys NFs confirm the high biocompatibility through cellular toxicity studies towards HFF cell lines. To better understand the antibacterial mechanism, flow cytometry and fluorescence imaging via DCFH-DA staining and GSH oxidation assays were performed to determine ROS-dependent and ROS-independent oxidative stress, respectively, *in-vitro*. Therefore, the results suggest that the antibacterial activity of $MoS_2$-cys NFs over both bacterial strains is due to the combined effect of membrane rupture, ROS-dependent, and ROS-independent oxidative stresses. Hence, our study indicates that the as-synthesized $MoS_2$-cys NFs exhibits remarkable antibacterial properties, high colloidal stability, and favourable biocompatibility towards mammalian cells.

## 2. Experimental section:

### 2.1. Materials:

Sodium molybdate dihydrate ($Na_2MoO_4 \cdot 2H_2O$, 99%), thiourea ($NH_2CSNH_2$, >99%), L-cysteine ($HSCH_2CH(NH_2)CO_2H$, >98.5%), were all purchased from Sigma-Aldrich. MTT (3-(4,5-dimethylthiazol-2-yl)-2,5-diphenyl tetrazolium bromide) reagent, Glutathione (GSH), and 5,5′-dithiobis(2-nitrobenzoic acid) (DTNB), and Luria-Bertani broth (LB) were acquired from Sisco Research Laboratories Pvt. Ltd. (SRL) India. Dimethyl sulfoxide (DMSO), Hydrogen Peroxide ($H_2O_2$), Phosphate-buffered saline (PBS, PH≈7.4), DCFH-DA (2′,7′-Dichlorofluorescein diacetate), were purchased from Sigma-Aldrich.

## 2.2. Synthesis of MoS$_2$-cys NFs:

The MoS$_2$-cys NFs were synthesized using a previously reported method with some modifications[24]. Briefly 0.01M of Na$_2$MoO$_4$·2H$_2$O was dissolved in 55 ml of Deionized (DI) water, followed by 45 minutes of stirring. Subsequently, 0.01M L-cysteine was introduced into the solution and stirred for another 45 minutes. Then, thiourea (2:1) was slowly added to the solution, and the mixture was again stirred for another 30 mins. Then, the obtained solution was transferred into a 100 ml Teflon-lined stainless-steel autoclave and heated at 200 °C for 24 hrs. After allowing the autoclave to cool, the suspension was washed several times with DI water and absolute ethanol, respectively. Finally, after vacuum drying, we obtained MoS$_2$-cys NFs. The schematic of the synthesis process is shown in Fig.1

## 2.3. Characterisation of MoS$_2$-cys NFs:

The X-ray diffraction (XRD) analysis was conducted using the X'Pert Pro high-resolution X-ray diffractometer (Philips PANalytical). The measurements were taken within the 2θ angle range of 10° to 70°, using Cu Kα radiation (λ=1.5406 Å). The size and morphology of the particles were determined using Field Emission-Scanning Electron Microscopy (FE-SEM, CARL ZEISS SUPRA 40, 20 kV) equipped with Energy Dispersive X-ray (EDX), and transmission electron microscopy (TEM, FEI-TECNAI G220S-Twin operated at 200 kV). X-ray photoelectron spectroscopy (XPS) measurements with a monochromatic Al-Kα source were recorded under high vacuum conditions using a PHI-5000 VERSA Probe II, Inc., Japan. Fourier Transform infrared (FTIR) spectroscopy measurements were done using Thermofisher Scientific Instruments, USA, Model: Nicolet 6700. The thermogravimetry analysis (TGA) was done using a PerkinElmer

Diamond TG/DTA thermal analyzer at a heating rate of 10 °C/min from 25 °C to 750 °C under an air atmosphere. The colloidal stability studies were done using a microplate reader (BioRad, iMark, Japan) by measuring optical density (OD) value at wavelength 595 nm and using a dynamic light scattering (DLS) experiment by measuring mean scattered light intensity from the sample at 90° with the help of Photocor complex, Photocor Ltd. Russia. The light source is a Ga-As diode laser at wavelength 653.8 nm and constant output power of 35 mW.

### 2.4. Assay for antibacterial activity of MoS$_2$-cys NFs:

An antimicrobial growth kinetics study examined the effect of as-synthesized MoS$_2$-cys NFs over bacterial growth. The two bacterial strains used in this study were gram-negative *E. coli* and gram-positive *S. aureus* bacteria. We incubated both bacterial cultures separately in fresh LB broth medium at 37 °C and 150 revolutions per minute for 16 hours. The bacterial inoculums were prepared (at $10^7$-$10^8$ CFU/ml), cultured in 96 well plates, and treated with increasing concentrations of MoS$_2$-cys NFs from 31.25 μg/ml to 250 μg/ml against both *E. coli* and *S. aureus* bacteria separately. Here, the inoculum with fresh LB medium and bacteria (non-treated) was considered the negative control. The effectiveness of the treatment was determined by establishing the growth curve of bacteria by measuring the OD value at 595 nm using a microplate reader (BioRad, iMark, Japan) every 30 minutes for 6 hours. The results were also given as the inhibition percentage of treated cell growth compared to the growth of negative control (not treated) cells. The following equation can estimate the percentage inhibition:

$$\% \text{ Inhibition} = \frac{\text{OD}_{\text{control}} - \text{OD}_{\text{treated}}}{\text{OD}_{\text{control}}} \times 100 \qquad (1)$$

## 2.5. Intracellular ROS detection Assay:

The amount of intracellular ROS in bacterial cells was determined by following a specific ROS detection assay using flow cytometry and fluorescence imaging using a fluorescent probe DCFH-DA[41]. In this study, *E. coli* and *S. aureus* were allowed to grow in an incubator overnight at 37 °C. The cells were then treated with $MoS_2$-cys NFs at concentrations of 62.5 µg/mL and 250 µg/mL at $10^7$-$10^8$ CFU and kept at 37 °C. An untreated sample was used as a negative control for each bacterial strain. After 6 hours of treatment, cell pellets were collected by centrifugation at 11,000 rpm for 10 min and washed thrice with PBS. The cells were then stained with 10 µM of DCFH-DA solutions and incubated at 37 °C for 30 min in the dark. The production of ROS in the cells in response to $MoS_2$-cys NFs was analyzed by imaging using fluorescence microscopy at 40X magnification (Nikon eclipse,Ts2R). Another set-up for flow cytometry analysis is prepared in the same way to quantify the level of cellular oxidative stress triggered by increasing dosages (62.5 and 250 µg/mL) of $MoS_2$-cys by measuring fluorescence intensity using flow cytometry (BD LSRFortessa™).

## 2.6. Ellman's assay: GSH oxidation test

The loss in GSH activity is well determined using Ellman's assay according to previously reported protocol [13]. For this study, $MoS_2$-cys NFs were diluted in 50 mM bicarbonate buffer (of pH 8.6) to a concentration of 31.25 µg/mL, 62.5 µg/mL, 125 µg/mL, 250 µg/mL, 500 µg/mL, and 1 mg/mL. With the dilutions, GSH (0.8 mM) was added to the microcentrifuge tube to initiate the oxidation process. $H_2O_2$ (1 mM) was a positive control, while GSH (0.8 mM) was the negative control. The mixture was subsequently covered with alumina foil to prevent light exposure and placed in a shaker at room temperature at 150 rpm for 2 and 4 hours, respectively. After incubation, 0.05 M Tris-HCl and 100 mM DTNB (Ellman's reagent) were added to the mixture. A syringe filter (0.22 µm membrane filter) was used to filter the MoS$_2$-cys NFs from the mixture. A 96-well plate was loaded with 100 µL aliquots of the filtered solution. The absorbance was then determined at 412 nm using the microplate spectrophotometer (BioRad, iMark, Japan). The equation estimates the percentage loss of GSH:

$$\% \text{ Loss of GSH } = \left(1 - \frac{\text{OD(treated with NFs)}}{\text{OD(untreated)}}\right) \times 100 \qquad (2)$$

### 2.7. Cellular toxicity studies:

The cytotoxicity of MoS$_2$-cys NFs was estimated on a normal human foreskin fibroblast (HFF) cell line, *in-vitro*. In a 96-well plate, 2.96 X 10$^6$ normal HFF cells were plated and allowed to attach in an incubator for 24 hours. After 24 hours, the cells were treated with a serum-free medium supplemented with different concentrations of MoS$_2$-cys NFs and incubated for 48 hours. After 48 hours of incubation, cells were assessed by the 3-(4,5-dimethylthiazol-2-yl)-2,5-diphenyl tetrazolium bromide (MTT, Sigma Aldrich, USA) assay protocol. The optical density was taken using a microplate spectrophotometer (BioRad, iMark, Japan) at 595 nm, and the results were analyzed. Cell viability was determined using the following formula:

$$\text{Cell viability} = \frac{\text{Absorbance of treated HFF cells}}{\text{Absorbance of the control cells}} \times 100$$

### 3. Results and Discussion:

#### 3.1. Synthesis and characterization of MoS$_2$-cys NFs:

The L-cysteine-capped MoS$_2$ NFs were successfully synthesized using a hydrothermal process. The reaction mechanism in Fig.1 shows that initially, during the crystallization process, MoS$_2$ NSs are formed. Then, the edge-to-edge, edge-to-surface, and few surface-to-surface attachments during the reaction led to the formation of the NFs[31].

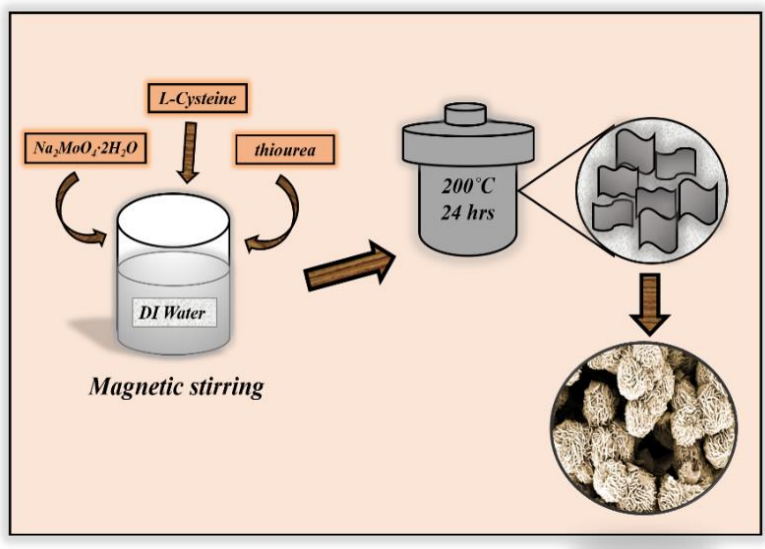

Figure 1. Schematic representation of the synthesis process and growth mechanism of MoS$_2$-cys NFs

The morphology and size of the as-obtained MoS$_2$-cys NFs were validated using FE-SEM and TEM. SEM image (Fig. 2a) shows the flower-like morphology of cysteine-capped MoS$_2$ nanostructures with plenty of edges. It can be observed that the NFs are composed of MoS$_2$ NSs arranged in the form of petals. The greater surface area and more edges in the NFs will help in more significant interaction of NFs with the bacteria and hence, may help increase the antibacterial activity[42]. The mean size of obtained NFs is 537± 12 nm, as calculated using the particle distribution curve in Fig. 2b.

The TEM images (Fig. 2c and S3) show that NFs are made up of few-layer MoS$_2$ NSs closely packed to form a flower-like structure. The EDX Spectrum (Fig.S1) and Elemental mapping analysis (Fig.S2) confirm the uniform distribution of Mo, S, C, N, and O among MoS$_2$-cys NFs.

The above results suggest that the prepared nanostructures could be MoS$_2$ NFs capped by L-cysteine.

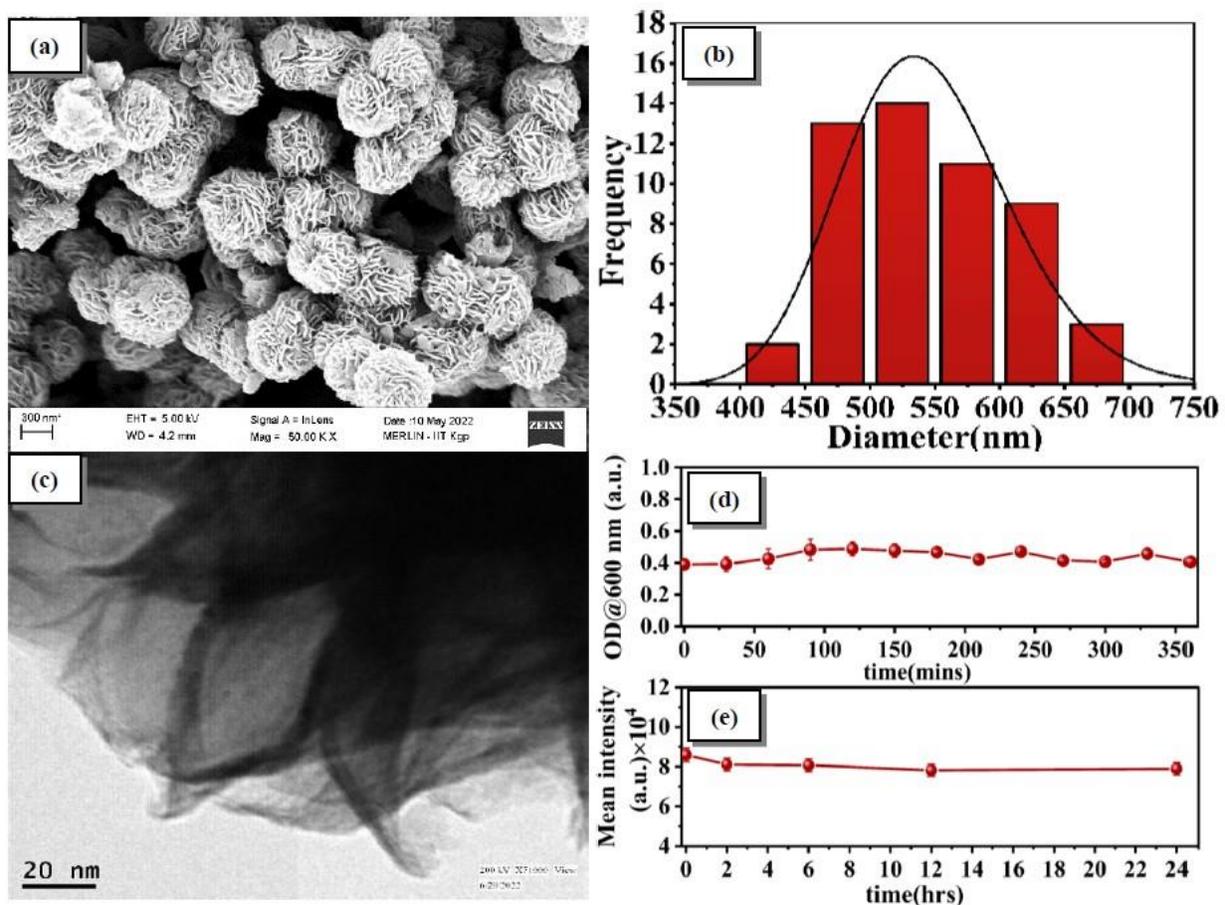

Figure 2. Morphology characterizations and colloidal stability of MoS$_2$-cys NFs, (a) SEM image of MoS$_2$-cys NFs, (b) corresponding size distribution, (c) TEM images of MoS$_2$-cys NFs, and (d & e) colloidal stability measurements of MoS$_2$-cys NFs using optical density values and mean scattered intensity, respectively with time

Further, Fig. 2d and 2e represent the colloidal stability studies of MoS$_2$-cys NFs using the OD values and the mean scattered intensity measurements, respectively. The previous studies found that aggregation of MoS$_2$ reduces its antibacterial effectiveness[13,43]. Hence, first using a microplate reader, OD values of the NFs at the concentration of 250 µg/ml are measured. The stability is investigated for the duration at which experimental studies for antibacterial activities are being performed. The increase in OD values usually indicates aggregation or poor stability of the complexes[44]. But, Fig. 2d shows that the OD value increases only nearly 2 % after 6 hours of incubation at 37°C. Therefore, the results suggest that the stability of the $MoS_2$-cys NFs is not affected during experimental conditions.

Moreover, the colloidal stability of nanoparticles is also assesed by measuring the mean scattered intensity from the NFs using DLS. The reduction in mean intensity indicates, aggregation of nanoparticles due to which they settle down. Fig. 2e shows only a 3 % decrease in intensity indicating minimal aggregation of nanoparticles even after 24 hours of incubation. These findings suggests the good colloidal stability of the $MoS_2$-cys NFs. However, it is known that the nanoparticles have a tendency to aggregate. But, their aggregation can be minimized with proper surface modification. Fig. S4 illustrates only 30 % of aggregation, even after 28 days of incubation. Hence, the above results reveals that the as-synthesized $MoS_2$-cys NFs are highly stable in aqueous media.

The possible crystal structure of as-prepared $MoS_2$-cys NFs was determined by analyzing the XRD pattern. From Fig.2, the diffraction peaks of the $MoS_2$-cys NFs correspond to (002), (100), (101), (006), (105), (110), and (008) lattice planes. These diffraction peaks could be indexed to the hexagonal structure of $MoS_2$ (JCPDS No. 37-1492). No other impurity peaks appeared in the XRD pattern, indicating that the formed $MoS_2$-cys structures are highly pure. The broadening of diffraction peaks suggests that the ultrathin dimensions of $MoS_2$ NSs in the NFs, which aligns with the observations from TEM images [24]. The crystallite size is found to be 2.02 nm, calculated using the standard Debye Scherrer formula.

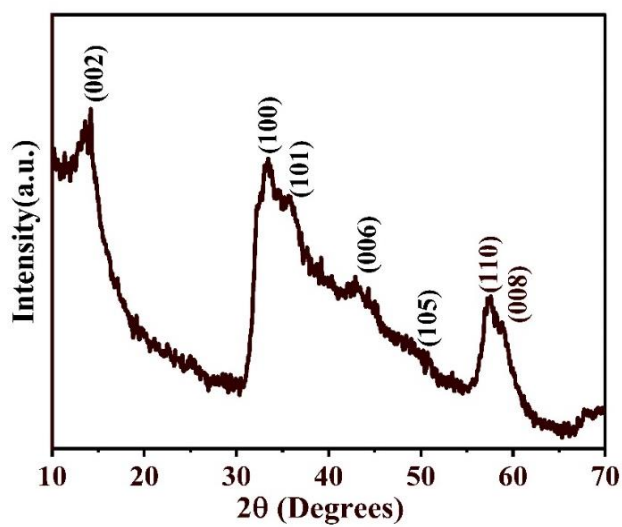

Figure 3. XRD spectra of MoS$_2$-cys NFs

The XPS was used to investigate the chemical composition and valence states of the synthesized MoS$_2$-cys NFs. The broad XPS spectrum is shown in Fig. 4a, which confirms the presence of Mo, S, C, and O in the synthesized sample. The high-resolution XPS spectrum of Mo and S is shown in Fig. 4b and c, respectively, as shown in the Mo 3d XPS spectrum of MoS$_2$-cys NFs, the two peaks at 229.2 eV and 232.4 eV can be assigned to Mo 3d$_{5/2}$ and Mo 3d$_{3/2}$ modes of 2H-MoS$_2$ in the Mo$^{4+}$ valence state[13,45].

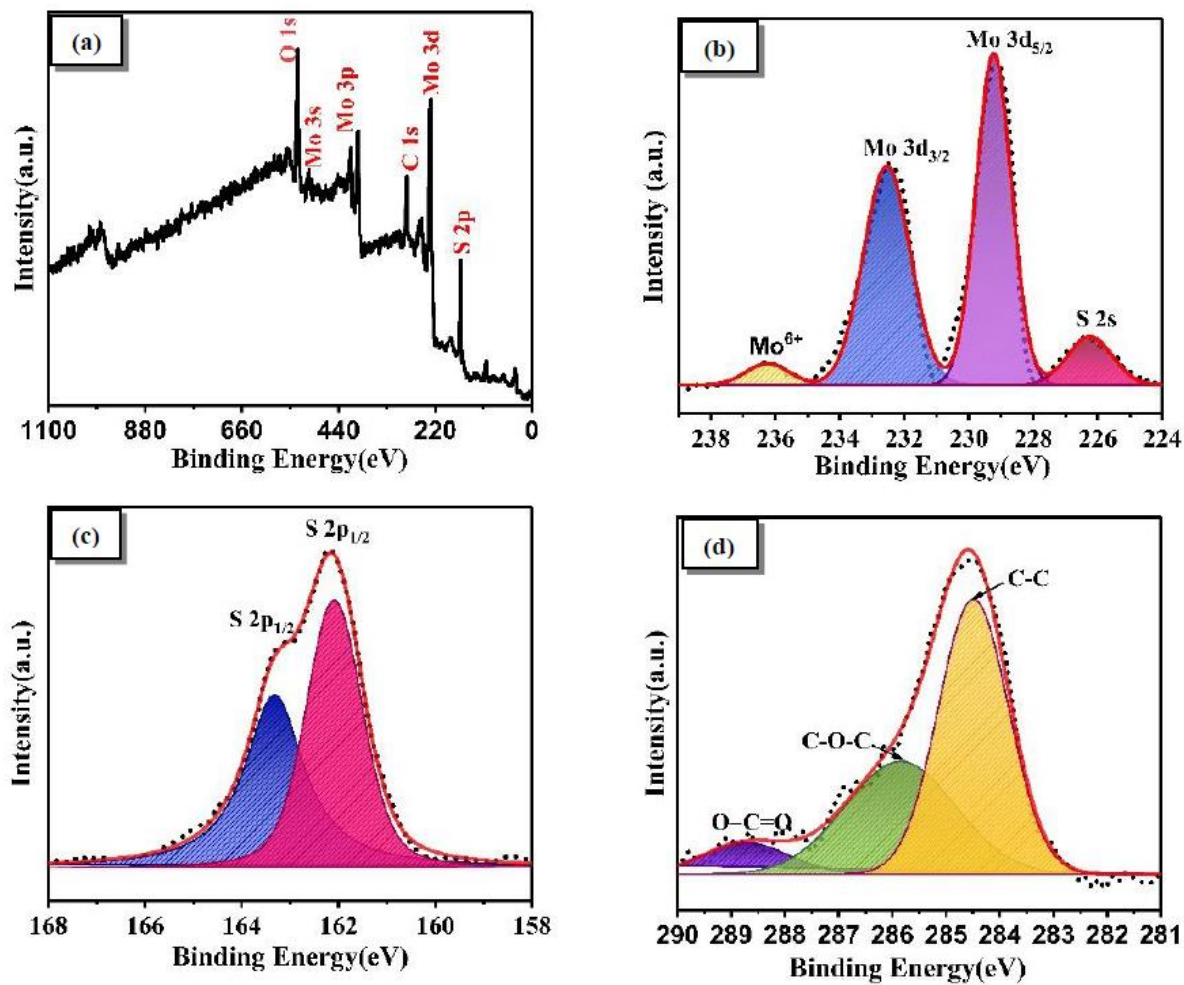

Figure 4. XPS spectra of MoS$_2$-cys NFs, (a) complete XPS scan of MoS$_2$-cys NFs (b) XPS spectra of Mo 3d, (c) XPS spectra of S 2p, and (d)XPS Spectra of C1s.

Additionally, there is a small peak at 236.1 eV, which corresponds to the Mo$^{6+}$ valence state of Mo. Further, one small peak at 226.3 eV corresponds to S 2s. Similarly, the characteristic peak at 162.1 eV and 163.2 eV corresponds to the S 2p$_{3/2}$ and S 2p$_{1/2}$ corresponding to the S$^{2-}$ valence state of the sulfur element[13,24]. These results indicate the formation of MoS$_2$. Fig. 3d shows the High-resolution XPS C1s spectrum where three peaks appeared at 284.5 eV (C-C), 285.85eV (C-O-C/C-OH), and 288.86 (O-C=O)[46,47] These peaks may be due to the presence of cysteine over the surface of MoS$_2$-cys NFs. Further, FTIR measurements were done to verify the presence of L-cysteine in MoS$_2$-cys NFs. FTIR spectrum of MoS$_2$-cys NFs and L-cysteine is shown in Fig. 5a. In FTIR spectra of L-cysteine, the main absorption bands at 1606 and 1389 cm$^{-1}$ can be assigned to asymmetric and symmetric stretching of COO$^-$, 2545 cm$^{-1}$ corresponding to S-H[20,48], and 2074 cm$^{-1}$ to C-H vibrations. Moreover, the band at 1525 cm$^{-1}$ corresponds to the N-H vibrations[20], 1054 cm$^{-1}$ to NH$_3$ rocking, and 1192 cm$^{-1}$ corresponds to CH$_2$ twisting[49]. The absorption band from 3200 to 3500 cm$^{-1}$ belongs to the O-H group. This spectrum matches well with that of standard amino acids[20]. Whereas in MoS$_2$-cys NFs, the spectrum clearly shows the distinctive peaks of Mo-O and Mo-S vibrations at around 600 and 481 cm$^{-1}$, respectively, corresponding to MoS$_2$, and the O-H stretching peak was found to be over 3330 to 3630 cm$^{-1}$ due to the absorption of atmospheric moisture by KBr[46]. Moreover, the signature peaks of L-cysteine in the MoS$_2$-cys NFs with reduced absorbance compared to L-cysteine are clearly visible. While the absorption band of COO$^-$ is shifted from 1606 to 1597 cm$^{-1}$ and from 1389 to 1397 cm$^{-1}$. Also, the peaks corresponding to C-H and S-H vibrations diminished in the NFs [20]. These results indicate the effective binding of L-cysteine with MoS$_2$.

The TGA curves of MoS$_2$-cys NFs and L-cysteine are shown in Fig. 5b. In the curve of MoS$_2$-cys NFs, the initial 2.85 % weight loss of NFs is due to evaporation of water from the complex and 18.78% weight loss from 100 °C to 750 °C in the TGA curve of NFs can be assigned to the thermal decomposition of L-cysteine[24]. Therefore, it shows that the amount of L-cysteine present in MoS$_2$-cys NFs is 18.78 %. The above results indicated the successful synthesis of the MoS$_2$-cys complex.

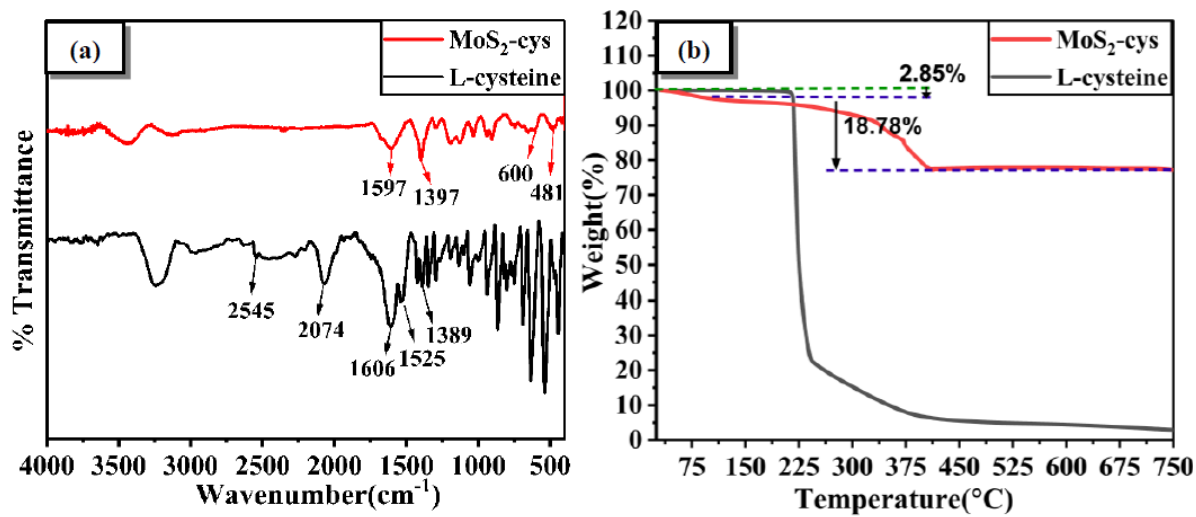

Figure 5. (a) FTIR spectra of MoS$_2$-cys NFs and (b)TGA curves of cysteine, and MoS$_2$-cys NFs

## 3.2. Antibacterial activity:

A bacterial growth kinetics study investigated the antibacterial activity of MoS$_2$-cys NFs against both Gram-negative *E. coli* and Gram-positive *S. aureus* bacteria. Both *E. coli* and *S. aureus* bacteria were exposed to MoS$_2$-cys NFs at varying concentrations. The growth curves demonstrate that the antibacterial activity of MoS$_2$-cys NFs against both the bacterial strains is concentration as well as time-dependent (Fig. 6a and 6b).

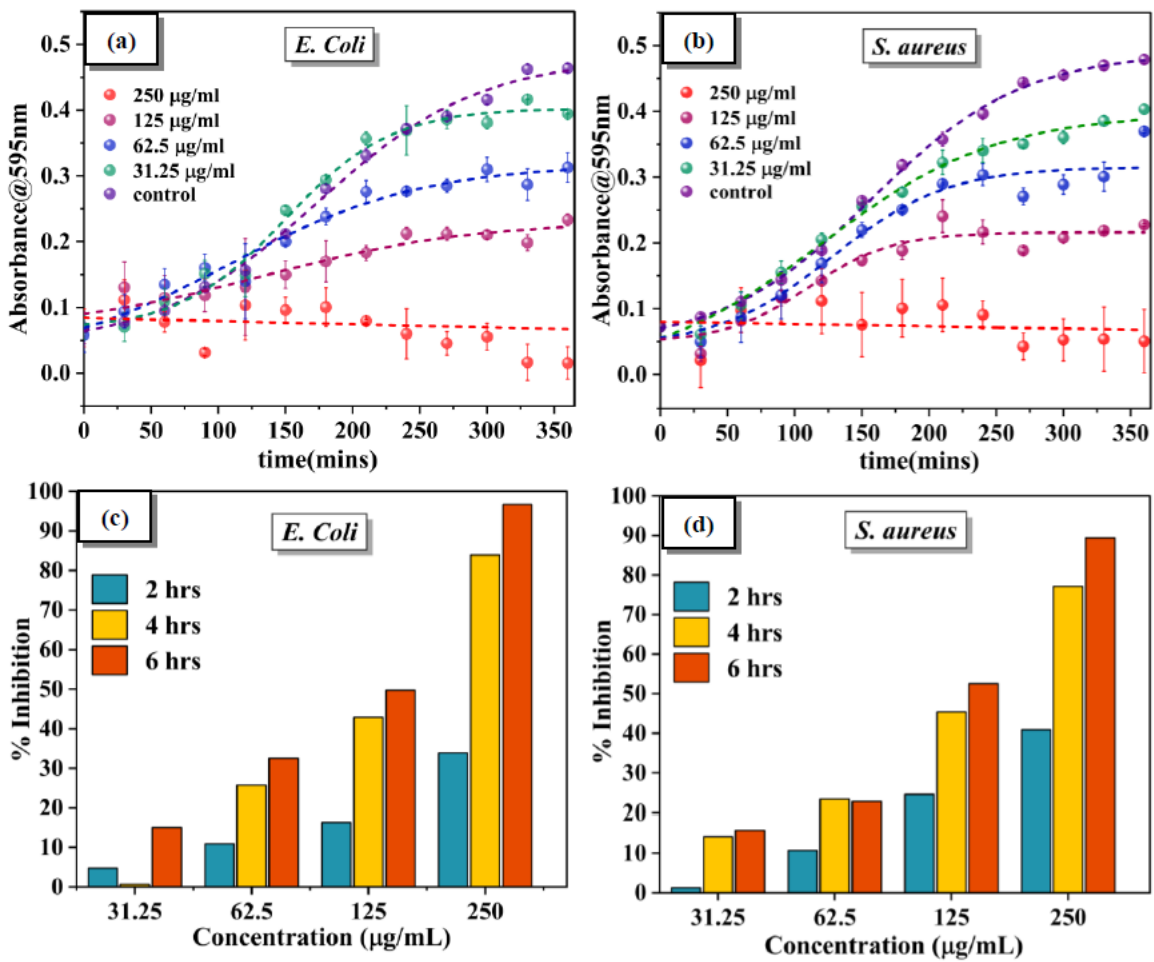

Figure 6. Antibacterial activity of MoS$_2$-cys NFs. Growth kinetics of (a) *E. coli* and (b) *S.aureus* treated with different concentrations of MoS$_2$-cys NFs ( n=3, ***$P<0.001$ is relative to control). % Inhibition of (c) *E.coli* and (d) *S.aureus* treated with different concentrations of MoS$_2$-cys NFs.

The percentage inhibition of treated cells relative to the control is plotted in Fig. 6 c and d to estimate the antibacterial activity of MoS$_2$-cys NFs. These results indicate that MoS$_2$-cys NFs have significantly reduced bacterial growth in both concentration- and time-dependent manner. After 6 hours of treatment, the MoS$_2$-cys NFs were found to inhibit the growth of *E. coli* by approximately 50 % at a concentration of 125 µg/mL which was increased to 97% at a concentration of 250 µg/mL. Moreover, the antibacterial efficiency of MoS$_2$-cys NFs against *E.coli* increased from 84% (after 4 hrs) to 97% (after 6 hrs) at a concentration of 250 µg/mL, indicating that this concentration of MoS$_2$-cys NFs is most effective in inhibiting the growth of Gram-negative bacterial cells (Fig. 6c).

Similar, antibacterial activities were also observed against the *S. aureus* bacteria, exhibiting a dependence on both concentration and time. However, at 250 µg/mL concentrations, *S. aureus* inhibition is nearly 75 % after 4 hours and almost 90 % after 6 hours. As a result, when compared to *S. aureus*, MoS$_2$-cys NFs were found to be more efficient at preventing the growth of *E. coli*. (Figure 6d.). A thick peptidoglycan layer surrounding gram-positive bacterial cells could explain the relatively less effectiveness of MoS$_2$-cys NFs in inhibiting the growth of *S. aureus* compared to *E. coli*[50].

| Study | Nanostructures | Incubation period (hours) | Light irradiation | Antibacterial activity % | | | | Colloidal stability | References |
|---|---|---|---|---|---|---|---|---|---|
| | | | | With light | | Without light | | | |
| | | | | *E. Coli* | *S. aureus* | *E. Coli* | *S. aureus* | | |
| Silver-Infused MoS$_2$ | 1. Cys- MoS$_2$ NSs (30 µg/mL) <br> 2. Ag$^+$-Cys-MoS$_2$ NSs (30 µg/mL) <br> 3. PDDA-Ag$^+$-Cys-MoS$_2$ NSs (15 µg/mL for *E. Coli* and 10 µg/mL for *S. aureus*) | 8 | - | - | - | 1. none <br><br> 2. 80% <br><br><br> 3. 100% | 1. none <br><br> 2. nearly 60% <br><br> 3. 100% | Not reported | 33 |
| PEG functionalized MoS$_2$ NFs | 1. PEG-MoS$_2$ NFs ( E.Coli at 100 and 1000 µg/mL) <br> 2. PEG-MoS$_2$ NFs+H$_2$O$_2$+light (*E. Coli* at 150 µg/mL) | 12 | 808 nm laser irradiation (for 10 min) | 1. - <br><br> 2. 99% | | 1.none and 15% respectively <br> 2. - | - | Not reported | 36 |
| PEG-MoS$_2$/rGO-SS nanoflakes | 1. PEG-MoS$_2$ <br> 2. PEG-MoS$_2$/rGO <br> 3. PEG-MoS$_2$/rGO-SS <br> 4. PEG-MoS$_2$/rGO-SS + light | 24 | 808 nm laser irradiation (for | <br><br><br><br> 4. 85% | <br><br><br><br> 4. 100% | 1. 20% <br> 2. 40% <br><br> 3. 60% <br><br> 4. - | 1. 40% <br> 2. 50% <br><br> 3. 60% <br><br> 4. - | Not reported | 37 |
| | (Both E. Coli and S. aureus at concentration 150 µg/mL for all above ) | | 30 min) | | | | | | |
| MoS$_2$ nanostructures | 1. MoS$_2$ NSs by ultrasonication 2. hydrothermally synthesized MoS$_2$ NFs . 3. MoS$_2$ NSs by lithium-ion intercalation (Conc. = 100 µg/mL) | 3 | White light LED irradiation( 18 W for 3hrs) | 1. 34% 2. 62% 3. 99% | - | 1. 15% 2. 18% 3. 36% | - | Antibacterial efficiency is decreased significantly by greater than 10% after 3$^{rd}$ cyclic antibacterial experimnets | 38 |
| MoS$_2$ NFs | 1. 2H- MoS$_2$ 2. 1T- MoS$_2$ (Concentration= 0.4 mg/L) | 3 | Simulated solar AM1.5 light (for 4 mins) | 1. 50% 2. 58% | | 1. no inhibition 2. 22% | | Not reported | 51 |
| MoS$_2$ QDs | 1. MoS$_2$ QDs 2. MoS$_2$ NSs 3. Bulk MoS$_2$ (Conc. = 50 µg/mL) | - | Simulated solar light (for 1 hour) | 1.~50% 2.~90% 3. none | 1. ~60% 2. ~90% 3. none | 1.~20% 2. - 3. - | 1. ~20% 2. - 3. - | Not reported | 52 |
| Amino acid functionalised MoS$_2$-QDS | 1. MoS$_2$ QDs (25 µg/mL) 2. phenylalanine functionalised MoS$_2$ QDs 3. Leucine functionalised MoS$_2$ QDs . | 14 | White light (30 min irradiation) | - | 1. . ~30% 2. ~99% (at 0.62 µg/mL) 3. ~99% (at 4 µg/mL) | - | 1. none 2. ~99% (at 7 µg/mL) 3. ~99% (at 1.24 µg/mL) | Not reported | 53 |
| **MoS$_2$-Cys NFs (Our work)** | MoS$_2$-Cys NFs: 1. 31.25 µg/mL 2. 62.5 µg/mL 3. 125 µg/mL 4. 250 µg/mL | 6 | - | - | - | 1. 15% 2. 32% 3. 50% 4. 97% | 1. 15% 2. 23% 3. 52% 4. 90% | Highly stable (~97% after 24 hours) | |

Table 1. Antibacterial efficiency of different MoS$_2$ nanostructures

Therefore, the above-mentioned findings in the table demonstrate the enhanced antibacterial effectiveness is either through external functionalisation, light exposure and/or combination with other nanoparticles. Thus, our results indicate that even without the aid of external probes (like visible light, NIR light, etc.) or loading of any antibiotics/drugs, doping , or combination with any other nanomaterials, the as-synthesized $MoS_2$-cys NFs alone exhibits remarkable efficiency towards bacterial inhibition. Also, due to the photothermal property of $MoS_2$ nanostructures as mentioned above, we may also expect the further increase in antibacterial potential of $MoS_2$-cys NFs.

**3.3. Mechanism of antibacterial action:**

It has been reported that $MoS_2$-NSs can embed into the lipid layer by forming dents on the surface, causing the rupture of the lipid membrane, cytoplasm leakage, and, eventually, bacterial death[54]. Moreover, the antibacterial mechanism of $MoS_2$-NSs by membrane depolarisation, bilayer disruption, oxidative stress, and metabolic inactivation has also been reported[40]. Since $MoS_2$-NFs consist of several $MoS_2$-NSs, similar antibacterial mechanisms with improved antibacterial efficiency are expected due to presence of many active edges. Apart from membrane-directed bacterial death, oxidative stress plays a major role in the antibacterial action of $MoS_2$-NSs.

Typically, oxidative stress can arise in different ways. One is ROS-dependent oxidative stress. ROS are widely known to cause cell apoptosis and contribute to the antibacterial activity of various nanomaterials. During cellular metabolism, ROS is constantly generated[55]. Even though bacteria have a defense system against ROS, excessive ROS will result in oxidative stress, destroying cellular components, including lipids, proteins, and nucleic acids, eventually killing the bacteria[56]. Other, ROS-independent oxidative stress, in which nanomaterials disturb a specific microbial process by disrupting or oxidizing a critical cellular structure or component without generating

ROS[13]. We followed established protocols to determine the oxidative stress-mediated antibacterial mechanism of the MoS$_2$-cys NFs.

### .3.1. Measurement of intracellular ROS:

First, we evaluated the possibility of ROS-generated oxidative stress by flow cytometry and fluorescence imaging using the DCFH-DA staining method. In the presence of ROS inside the cells, the non-fluorescent DCFH-DA dye is oxidized to form a green fluorescent molecule called dichlorofluorescein (DCF), which acts as the indicator of ROS[41]. Both the bacterial strains, *E. coli* and *S. aureus* were treated with different concentrations (0, 62.5, and 250 µg/mL) of MoS$_2$-cys NFs.

The results of flow cytometry-based ROS measurement clearly show a concentration-dependent ROS generation in both *E. coli* (Fig. 7a and S5) and *S. aureus* (Fig. 7b and S6) bacteria. In the case of *E. coli*, ROS generation is 52.5 % and 6 % after treatment with 250 µg/mL and 62.5 µg/mL concentrations of MoS$_2$-cys NFs, respectively, while the control only shows 0.7 % of ROS generation. While in the case of *S. aureus*, the ROS generation is 17.9 %, 2 %, and 0.54 % at the concentration of 250 µg/mL, 62.5 µg/mL, and 0 µg/mL(non-treated or control), respectively. The results also demonstrate that more ROS is generated in *E. coli* compared to *S. aureus*, which is consistent with the bacterial growth kinetics assay.

Further, fluorescence microscopy imaging using DCFH-DA staining was done to confirm the generation of ROS upon treatment with MoS$_2$-cys NFs. The intracellular ROS level is indicated by green fluorescence intensity[57]. The results show that both *E. coli* (Fig. 7(c-e)) and *S. aureus* (Fig. 7(f-h)) show enhanced green fluorescence and, thus, improved ROS production upon treatment with MoS$_2$-cys NFs as compared to their dark controls.

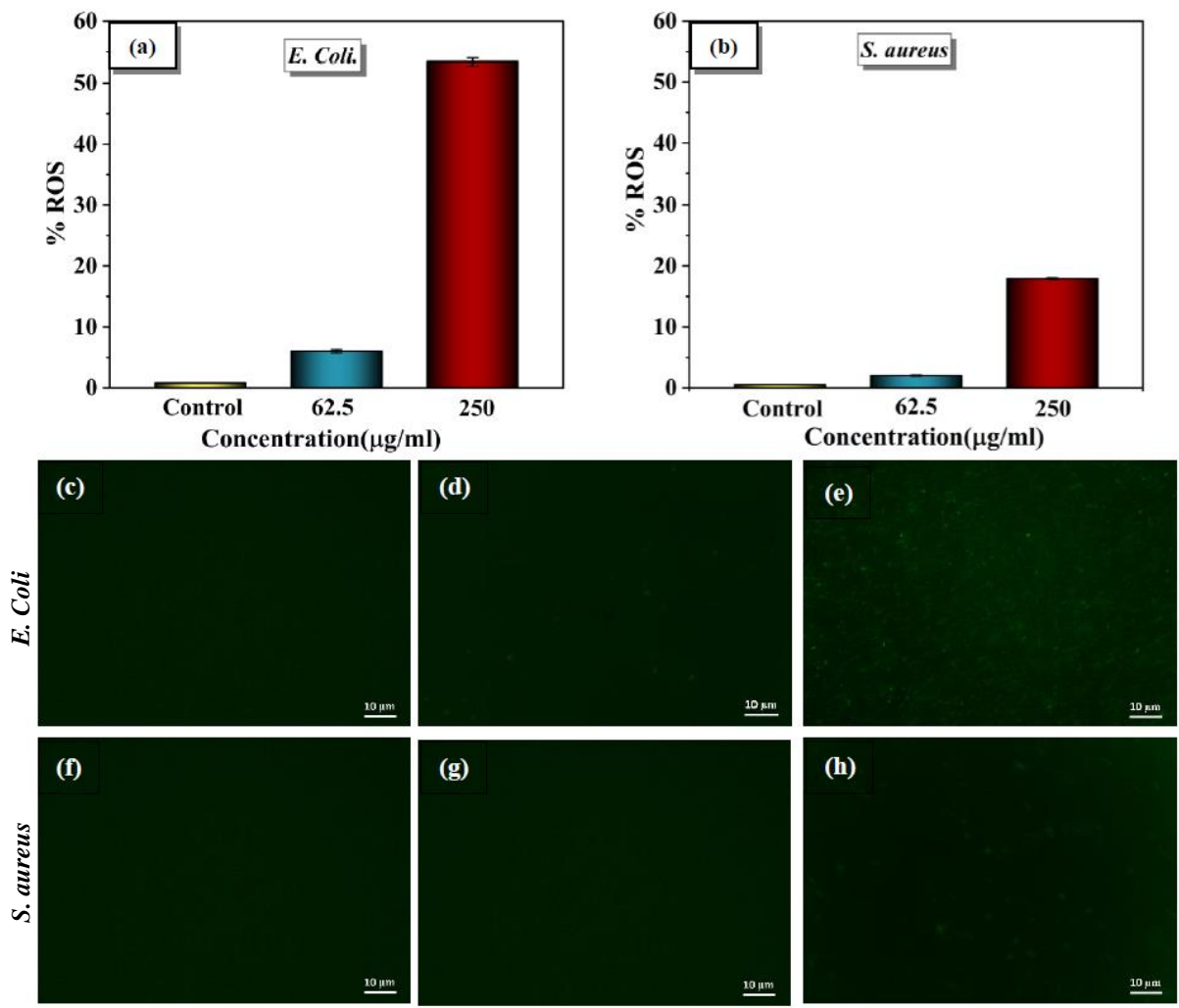

Figure 7. ROS generation at different doses of MoS$_2$-cys NFs. % ROS generation in (a)*E. Coli*, and (b) *S. aureus* after treatment with different concentrations of MoS$_2$-cys NFs (n=3, ***$P < 0.001$ relative to control). Fluorescence microscopy images of *E. coli* (c-e), and *S. aureus* (f-h) after exposure to 0 µg/mL(control) (c, f) ,62.5 µg/mL (d, g), and 250 µg/mL (e, h) of MoS$_2$-cys NFs.(scale bar: 10 µm, Magnification: 40X).

Moreover, the fluorescent images show that the fluorescence intensity is dose-dependent, suggesting the dose-dependent ROS generation in both bacterial strains. Also, compared to *E. coli*, the ROS generation is less in *S. aureus*. These results are consistent with the flow cytometry analysis as well as the bacterial growth kinetics study. The above results show ROS generation is a key factor in bacterial growth inhibition.

**3.3.2. Loss in GSH activity:**

Next, we investigated the possibility of ROS-independent oxidative stress using *in-vitro* GSH oxidation using Ellman's assay. GSH, a tripeptide that contains thiols, is present in the cells in large amounts in its reduced form, which spontaneously oxidizes upon exposure to molecular oxygen[58]. As an antioxidant, it protects cells from oxidative stress. Excess ROS within cells causes oxidative stress and disrupts the equilibrium between pro-oxidants and antioxidants, oxidizing GSH into glutathione disulfide (GSSG). The oxidation of GSH to GSSG leads to a loss in GSH activity.

Consequently, GSSG cannot protect cells from oxidative damage, resulting in the oxidation of vital cellular components such as proteins, nucleic acids, lipids, etc., ultimately leading to the death of the cell[50]. Ellman's assay can quantify the oxidative stress caused by GSH activity loss. In GSH oxidation experiments, bicarbonate buffer (50.0 mM at pH 8.6) without $MoS_2$-cys NFs and $H_2O_2$ (1.0 mM) were used as negative and positive controls, respectively.

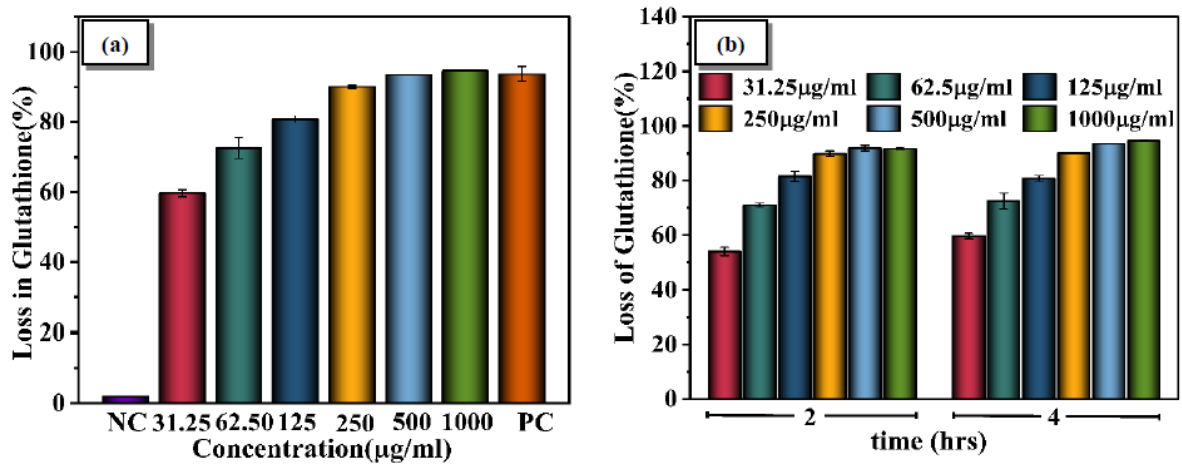

Figure 8. Oxidation of GSH by MoS$_2$-cys NFs. (a) Percentage loss of GSH after incubation with MoS$_2$-cys NFs dispersions of different concentrations for 4 h. H$_2$O$_2$ and bicarbonate buffer without NFs were used as positive and negative controls, respectively. (b) Time and concentration-dependent percentage loss of GSH. (n=3, ***$P < 0.001$).

The objective of the negative control was to show that the incubation conditions did not cause GSH oxidation. The oxidation capacity of the MoS$_2$-cys NFs toward GSH showed a clear dependence on their concentration but no significant difference with incubation time. At the concentration of 31.25 µg/mL, the loss of GSH activity is nearly 58% which increased to 89% at the concentration of 250 µg/mL and almost 90% at 500 µg/mL and 1000 µg/mL. The results show that when the concentration is increased beyond 250 µg/mL, there is no significant change in the loss of GSH (Figure 8a). Therefore, 250 µg/mL is the minimum concentration most effectively reducing the GSH activity. The above results agree well with the growth kinetics study of both bacteria. Also, from Fig. 8b, it is visible that there is no significant loss in GSH with time, suggesting that 2 hrs incubation time is enough to oxidize GSH. The strong oxidation of GSH by the MoS$_2$-cys NFs supports that the NFs can oxidize thiols or other cellular components, consequently killing the bacteria by interfering with their defense mechanism. Therefore, it is clear from the above results that the MoS$_2$-cys NFs can generate significant ROS-independent oxidative stress. Thus, from the above results, we see that the antibacterial mechanism of MoS$_2$-cys NFs is due to the combined effect of ROS-independent and ROS-dependent oxidative stress. Thus, from the above results and report findings, we propose that the large interaction sites over the surface of NFs and the thin dimension of NSs in the form of petals are capable of generating oxidative stresses and membrane destruction, leading to the death of the bacteria.

### 3.4. Cellular toxicity studies:

Fig. 9 shows the MTT assay evaluated cellular toxicity of $MoS_2$-cys NFs against the HFF cell line. The HFF cells were treated with different concentrations of $MoS_2$-cys NFs. Analysis of the results showed that after 48 hours of treatment with $MoS_2$-cys NFs, there was a minimal/non-significant decrease in cell viability at doses up to 250 µg/mL. It has been observed that, on average, 90% of cells are still viable after treatment. These results suggest that $MoS_2$-cys NFs cause negligible toxicity to eukaryotic cells. This finding implies that the nanoflower's safety and cytotoxicity profiles are within acceptable limits under normal *in-vitro* experimental conditions. It could be used as a non-toxic material for further research.

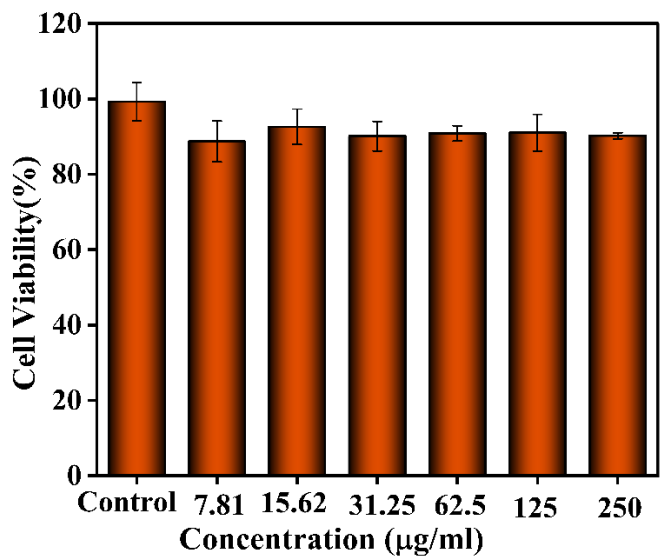

Figure 9. In-vitro cytotoxicity study of $MoS_2$-cys NFs. % Cell viability was assessed by MTT assay with varying concentrations. The cell viability of control cells was taken as 100%. The data shown are the mean ± SD (n=3).

## 4. Conclusions:

MoS$_2$-cys NFs were successfully synthesized using the simple hydrothermal method. Various characterizing techniques confirm the formation of cysteine-capped MoS$_2$ NFs. XRD confirmed the formation of 2H-phase and purity of MoS$_2$-cys NFs, and FE-SEM and TEM confirmed the size and flower-like morphology. The XPS also confirms the formation of MoS$_2$ and the surface modification with L-cysteine. FTIR further confirms the presence of L-cysteine, and DTA-TGA analysis estimates 18% L-cysteine over the surface. Moreover, the as-prepared NFs have also demonstrated very high colloidal stability in aqueous solutions throughout the experimental conditions with only nearly 3% aggregation in 24 hrs. Further, the antibacterial studies using the broth dilution method against both gram-negative and gram-positive bacterial strains revealed the high potential of these NFs in inhibiting their growth in a concentration and time-dependent manner. The incubation of 250 μg/mL of MoS$_2$-cys NFs with *E. coli* and *S. aureus*, respectively, for 6 hrs have shown more than 90% inhibition of both bacterial strains, confirming their potency as antibacterial agents. Moreover, the antibacterial mechanisms insights suggest that the presence of more interaction sites and thin dimension of nanosheets in MoS$_2$-cys NFs results in enhanced antibacterial activity. This enhancement is due to combined effect of membrane damage, ROS-dependent and, ROS-independent oxidative stresses. Furthermore, toxicity studies of MoS$_2$-cys NFs confirm the high biocompatibility of as-prepared NFs.

The antibacterial activity of non-toxic MoS$_2$-cys NFs suggests their future biomedical applications, such as wound healing, water disinfection, antibacterial coatings, and in textile and food industries. Additionally, due to the inherent photothermal property of MoS$_2$, our future study aims to investigate the potential enhancement of antibacterial efficiency of MoS$_2$-cys NFs upon light exposure. Moreover, due to the high biocompatibility and colloidal stability of the MoS$_2$-cys NFs, they may be used as a drug carrier in treating diseases like, cancer. Apart from biomedical applications, due to the greater surface area and active edges, the as-prepared nanostructures could be used in catalysis, sensing, supercapacitors, solar cells, hydrogen evolution applications, etc.

**Conflicts of interest:**

The authors declare no conflicts of interest.

**Acknowledgments:**

RK acknowledges the Ministry of Human Resource and Development, India, for the fellowship. ANG acknowledges the financial support from the Department of Science and Technology, India, grant no. CRG/2019/001684. We thank the central research facility for the XRD, FE-SEM, TEM, XPS, FTIR, and TGA measurements. The authors highly acknowledge the support provided by the SERB Department of Science and Technology, Govt. of India (JCB/2019/000008), and Indian Council of Medical Research (ICMR) (Sanction Letter No: 5/13/53/2020-NCD-III

(53) Mondal, A.; De, M. Amino Acid-Functionalized MoS 2 Quantum Dots for Selective Antibacterial Activity. *Cite This: ACS Appl. Nano Mater* **2021**. https://doi.org/10.1021/acsanm.1c03243.

(54) Dong, S. M.; Li, J.; Liu, L.; Wu, R.; Ou, X.; Tian, R.; Zhang, J.; Jin, H.; Dong, M. Membrane Destruction and Phospholipid Extraction by Using Two-Dimensional MoS 2 Nanosheets †. *Nanoscale* **2018**, *10*. https://doi.org/10.1039/c8nr04207a.

(55) Simon, H.-U.; Haj-Yehia, A.; Levi-Schaffer, F. Role of Reactive Oxygen Species (ROS) in Apoptosis Induction. *Apoptosis* **2000**, *5*, 415–418.

(56) Thannickal, V. J.; Fanburg, B. L. Downloaded from Journals.Physiology.Org/Journal/Ajplung. *Am J Physiol Lung Cell Mol Physiol* **2000**, *279*, 1005–1028.

(57) Xiong, Z.; Zhang, X.; Zhang, S.; Lei, L.; Ma, W.; Li, D.; Wang, W.; Zhao, Q.; Xing, B. Bacterial Toxicity of Exfoliated Black Phosphorus Nanosheets. *Ecotoxicol Environ Saf* **2018**, *161*, 507–514. https://doi.org/10.1016/J.ECOENV.2018.06.008.

(58) Fahey, R. C.; Brown, W. C.; Adams, W. B.; Worsham, M. B. Occurrence of Glutathione in Bacteria. *J Bacteriol* **1978**, *133* (3), 1126–1129. https://doi.org/10.1128/JB.133.3.1126-1129.1978.

**Graphical Abstract**

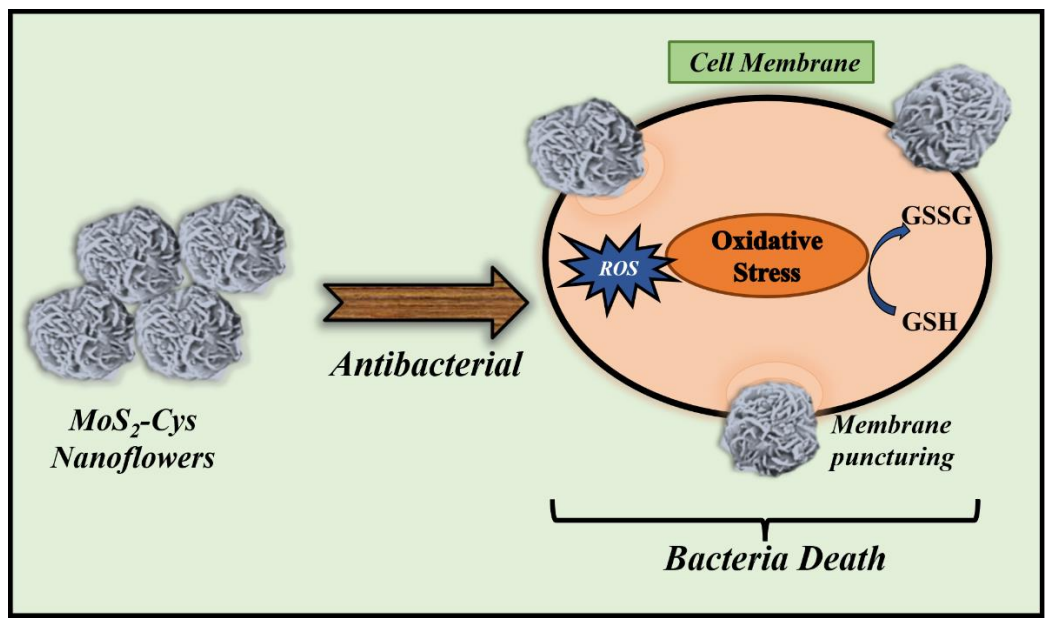